# Direct observation of competition between superconductivity and charge density wave order in YBa$_2$Cu$_3$O$_y$

J. Chang[1,2], E. Blackburn[3], A. T. Holmes[3], N. B. Christensen[4], J. Larsen[4,5], J. Mesot[1,2], Ruixing Liang[6,7], D. A. Bonn[6,7], W. N. Hardy[6,7], A. Watenphul[8], M. v. Zimmermann[8], E. M. Forgan[3], S. M. Hayden[9]

[1]*Institut de la matière complexe, Ecole Polytechnique Fedérale de Lausanne (EPFL), CH-1015 Lausanne, Switzerland.* [2]*Paul Scherrer Institut, Swiss Light Source, CH-5232 Villigen PSI, Switzerland.* [3]*School of Physics and Astronomy, University of Birmingham, Birmingham B15 2TT, United Kingdom.* [4]*Department of Physics, Technical University of Denmark, DK-2800 Kongens Lyngby, Denmark.* [5]*Laboratory for Neutron Scattering, Paul Scherrer Institut, CH-5232 Villigen PSI, Switzerland.* [6]*Department of Physics & Astronomy, University of British Columbia, Vancouver, Canada.* [7]*Canadian Institute for Advanced Research, Toronto, Canada.* [8]*Hamburger Synchrotronstrahlungslabor (HASYLAB) at Deutsches Elektron-synchrotron (DESY), 22603 Hamburg, Germany.* [9]*H. H. Wills Physics Laboratory, University of Bristol, Bristol, BS8 1TL, United Kingdom.*

**Superconductivity often emerges in the proximity of, or in competition with, symmetry breaking ground states such as antiferromagnetism or charge density waves (CDW)[1-5]. A number of materials in the cuprate family, which includes the high-transition-temperature (high-$T_c$) superconductors, show spin and charge density wave order[5-7]. Thus a fundamental question is to what extent these ordered states exist for compositions close to optimal for superconductivity. Here we use high-energy x-ray diffraction to show that a CDW develops at zero field in the normal state of superconducting YBa$_2$Cu$_3$O$_{6.67}$ ($T_c$ = 67 K). Below $T_c$, the application of a magnetic field suppresses superconductivity and enhances the CDW. Hence, the CDW and superconductivity are competing orders in this typical high-$T_c$ superconductor, and high-$T_c$ superconductivity can form from a pre-existing CDW state. Our results explain observations of small Fermi surface pockets[8], negative Hall and Seebeck effect[9,10] and the "$T_c$ plateau"[11] in this material when underdoped.**

Charge density waves in solids are periodic modulations of conduction electron density. They are often present in low dimensional systems such as NbSe$_2$[4]. Certain cuprate materials such as La$_{2-x-y}$Nd$_y$Sr$_x$CuO$_4$ (Nd-LSCO) and La$_{2-x}$Ba$_x$CuO$_4$, (LBCO) also show charge modulations that suppress superconductivity near x=1/8[6,7]. In some cases, these are believed to be unidirectional in the CuO$_2$ plane, and have been dubbed 'stripes'[2,3]. There is now a mounting body of indirect evidence that charge and/or spin density waves may be present at high magnetic fields in samples with high $T_c$: quantum oscillation experiments on underdoped YBa$_2$Cu$_3$O$_y$ (YBCO) have revealed the existence of at least one small Fermi surface pocket[8,9] which may be created by a charge modulation[10]. More recently, NMR studies have shown a magnetic-field-induced splitting of the Cu2F lines of YBCO[12]. An important issue is the extent to which the tendency towards charge order exists in high-$T_c$ superconductors[2,3].

Here we report a hard (100 keV) X-ray diffraction study, in magnetic fields up to 17 T, of a de-twinned single crystal of YBa$_2$Cu$_3$O$_{6.67}$ (with ortho-VIII oxygen ordering[11,13], $T_c$ = 67 K and $p$ = 0.12 where $p$ is the hole concentration per planar Cu). We find that a CDW forms in the normal state below $T_{CDW} \approx 135$ K. The charge modulation has two fundamental wavevectors $\mathbf{q}_{CDW} = \mathbf{q}_1 = (\delta_1, 0, 0.5)$ and $\mathbf{q}_2 = (0, \delta_2, 0.5)$, where $\delta_1 \approx 0.3045(2)$ and $\delta_2 \approx 0.3146(7)$. These give satellites of the parent crystal Bragg peaks at positions such as $Q=(2\pm\delta_1, 0, 0.5)$. Although the satellite intensities have a strong temperature and magnetic field dependence, the CDW is not field-induced and is unaffected by field in the normal state. Below $T_c$ it competes with superconductivity, and a decrease of the CDW amplitude in zero field becomes an increase when superconductivity is suppressed by field.

Figure 1a,g shows scans through the $(2-\delta_1, 0, 0.5)$ and $(0, 2-\delta_2, 0.5)$ positions at $T = 2K$. Related peaks were observed at $(2+\delta_1, 0, 0.5)$ and $(4-\delta_1, 0, 0.5)$ (see Supplementary Figure S3). The incommensurate peaks are not detected above 150K (Fig. 1c). From the peak width we estimate that the modulation has an in-plane correlation length $\xi_a \approx 98 \pm 5$ Å (at 2 K and 17 T). The existence of four similar in-plane modulations $(\pm\delta_1, 0)$ and $(0, \pm\delta_2)$ indicates that the modulation is associated with the (nearly square) $CuO_2$ planes rather than the CuO chains. The present experiment cannot distinguish between 1-*q* and 2-*q* structures i.e. we cannot tell directly whether modulations along *a* and *b* co-exist in space or occur in different domains of the crystal. However, Bragg peaks from the two CDW components have similar intensities and widths (Fig. 1b, g) despite the orthorhombic crystal structure, which breaks the symmetry between them. This suggests that $q_1$ and $q_2$ are coupled, leading to the co-existence of multiple wavevectors as seen in other CDW systems such as $NbSe_2$[4]. The scan along the *c** direction in Fig. 1d has broad peaks close to $\ell=\pm0.5$ r.l.u., indicating that the CDW is weakly correlated along the *c* direction with a correlation length $\xi_c$ of approximately 0.6 lattice units.

In zero field, the intensity of the CDW Bragg peak (Fig. 2) grows on cooling to $T_c$ below which it is partially suppressed. A magnetic field applied along the *c* direction has no effect for $T>T_c$. Below $T_c$ it causes an increase of the intensity of the CDW signal (Figs. 1a, 2). At $T=2$ K, the intensity grows with applied magnetic field (Fig. 2b) and shows no signs of saturation up to 17 Tesla. The magnetic field also makes the CDW more long range ordered (Fig. 2c). In zero magnetic field, the *q*-width varies little with temperature. However, below $T_c$ in a field, the CDW order not only becomes stronger, but also becomes more coherent, down to a temperature $T_{cusp}$ below which the intensity starts to decrease (Figs. 2, 4). All of this is clear evidence for competition between CDW and superconducting orders.

Non-resonant X-ray diffraction is sensitive to modulations of charge density and magnetic moments. In our case, the expected magnetic cross-section is several orders of magnitude smaller than our observed signal, which must therefore be due to charge scattering. NMR measurements on a sample of the same composition as ours[12] indicate that the CDW is not accompanied by magnetic order. Charge density modulations in solids will always involve both a modulation of the electronic charge and a periodic displacement of the atomic positions[14]. We are more sensitive to the atomic displacements than to the charge modulation because ions with large numbers of electrons (as in YBCO) dominate the scattering. The scattering arises only from atomic displacements parallel to the total scattering vector *Q*. The comparatively small contribution to *Q* along *c** from $\ell = 0.5$ r.l.u. means that our signal for a $(h, 0, 0.5)$ peak is dominated by displacements parallel to **a**. (There will also be displacements parallel to **c** but we are insensitive to them in our present scattering geometry). Our data indicate that the incommensurate peaks are much stronger if they are satellites of strong Bragg peaks of the form $(\tau = (2n, 0, 0))$ at positions such as $\tau\pm q_1$. This indicates that the satellites are caused by a modulation of the parent crystal structure. The fact that the scattering is peaked at $\ell = \pm 0.5$ r.l.u. means that neighbouring bilayers are modulated in antiphase. The two simplest structures (Fig.3a, b) compatible with our data involve the neighbouring $CuO_2$ planes in the bilayer being displaced in the same (bilayer-centred) or opposite (chain-centred) directions resulting in the maximum amplitude of the modulation being on the $CuO_2$ planes and CuO chains respectively. In their 2-$q$ form, these structures lead to the in-plane "checkerboard" pattern shown in Fig. 3(c). STM studies of other underdoped cuprates[15] and of field-induced CDW correlations in vortex cores[16] also support the tendency towards checkerboard formation[17]. Our observation of a CDW may be related to studies of phonon anomalies[18], which suggest that in YBCO near 1/8 there are anomalies in the underlying charge susceptibility for $q \approx (0,0.3)$.

It is interesting to compare the underlying Fermi surface (FS) (Fig. 3d) of YBCO[19,20] (see Supplementary Information) with $q_{CDW}$ (Fig. 3d). States connected by the CDW ordering wavevector become coupled because of the additional potential introduced. The electronic states most obviously connected by $q_{CDW}$ are the bonding bands at and near the zone boundary, which lie in the anti-nodal region of the superconducting gap, where the pseudogap is also maximized. Incommensurate CDW order will cause the electronic band structure to fold and the FS will be reconstructed creating small pockets[21-22]. This qualitatively explains the changes in anisotropy observed in transport measurements[24,26] and the observation of low-frequency quantum oscillations[8].

The observation of a CDW in $YBa_2Cu_3O_{6.67}$ contrasts with the materials Nd-LSCO and LBCO, where rotations of the $CuO_6$ octahedra favour stripe-like charge and magnetic ordering[6,7]. In our case, NMR measurements[12] indicate that there is no magnetic order. They also suggest that CDW order only appears at $T \approx T_c$ for $H > 10$ T; it is possible that the CDW is only frozen on the NMR timescale at high fields and lower temperatures.

Our results have important implications for the phenomenology and phase diagram of the cuprates (Fig. 4). One of the defining properties of underdoped cuprates such as ortho-VIII YBCO is the pseudogap. This develops at the "crossover" temperature $T^*$ (for $YBa_2Cu_3O_{6.67}$ $T^* \approx 220$ K), where there is a suppression of low energy electronic states, and rotational anisotropy appears in various physical properties, such as the Nernst effect[23]. The CDW reported here develops at $T_{CDW} \approx 135$ K, inside the pseudogap state. It is interesting to note that $T_{CDW}$ corresponds approximately with $T_H$ (Fig. 4), the temperature at which Hall effect measurements suggest that Fermi surface reconstruction begins[24] and where anisotropy in the Nernst effect is suppressed[25]. Both of these effects may be caused by the CDW. We can make further connections with the cuprate phase diagram in a field. From Figure 2a,c we can identify the temperature $T_{cusp}(H)$ where the suppression of the CDW begins. This temperature (Fig. 4) appears to correspond to the temperature $T_n(H)$ at which superconductivity first forms and the vortex liquid phase appears as identified from transport measurements. Thus, the competition starts in the regime of flux flow and superconducting *fluctuations*[26,27] (rather than zero-resistance superconductivity). A simple Landau theory [see the supplementary information] shows that $T_c$ will be suppressed below the value it would have in the absence of the CDW[28,29]. We speculate that this is reflected in the shape of the superconducting dome[11] (Fig. 4).

In conclusion, we have observed a clear coupling between the superconducting and CDW order parameters. On cooling below the superconducting $T_c$, the CDW order parameter decreases. Application of a magnetic field at low temperatures suppresses superconductivity, and a concomitant increase in the CDW order parameter confirms the strong coupling of order parameters. High-$T_c$ superconductivity in ortho-VIII ($p \approx 1/8$) YBCO develops out of a pre-existing and competing CDW ordered state, which exists for $T \gg T_c$ and most likely persists to high fields $H \gg H_{c2}$.

**Acknowledgements:**
We thank B. L. Gyorffy, M. W. Long, J. A. Wilson and A. J. Schofield for discussions and R. Nowak, G. R. Walsh, and J. Blume for technical assistance. This work was supported by the EPSRC (grant numbers EP/G027161/1 and EP/J015423/1), the Wolfson Foundation, the Royal Society and the Danish Agency for Science, Technology and Innovation under DANSCATT. J.C., N.B.C. and J.M. are grateful to L. Braicovich, G. Ghiringhelli, B. Keimer and M. Le Tacon for communicating their results[31] to them after this experiment was completed.

**Author Contributions:**
D.A.B., W.N.H, and R. L., prepared the samples. E.B., J.C., N.B.C., E.M.F., S.M.H., M. v. Z., A. W. conceived and planned the experiment. E.B., J.C., N.B.C., E.M.F., S.M.H., A.T.H., J. L., M. v. Z. carried out the experiment. J.C., E.M.F and S.M.H. carried out data analysis and modelling. E.B, J.C., N.B.C., E.M.F and S.M.H. wrote the paper. J. M. was responsible for research direction and planning at PSI. All authors discussed the results and commented on the manuscript.

**Author Information:**
Correspondence and requests for materials should be addressed to J.C. (johan.chang@epfl.ch).

## REFERENCES:

[1] Mathur, N. D., *et al*., Magnetically mediated superconductivity in heavy fermion compounds. *Nature* **394**, 39-43 (1998).

[2] Kivelson, S. A., *et al*., How to detect fluctuating stripes in the high-temperature superconductors. *Rev*. *Mod*. *Phys*. **75**, 1201-1241 (2003).

[3] Vojta, M., Lattice symmetry breaking in cuprate superconductors: stripes, nematics, and superconductivity. *Advances in Physics* **58**, 699-820 (2009).

[4] Moncton, D. E., Axe, J. D., and DiSalvo, F. J., Neutron scattering study of the charge-density wave transitions in 2H-TaSe$_2$ and 2H-NbSe$_2$. *Phys. Rev. B* **16**, 801-819 (1977).

[5] Demler, E., Sachdev, S., and Zhang, Y., Spin-Ordering Quantum Transitions of Superconductors in a Magnetic Field. *Phys. Rev. Lett.* **87**, 067202 (2001).

[6] Tranquada, J. M., *et al*., Evidence for stripe correlations of spins and holes in copper oxide superconductors. *Nature* **375**, 561 (1995).

[7] Fujita, M., Goka, H., Yamada, K., Tranquada, J. M. &Regnault, L. P., Stripe order, depinning, and fluctuations in La$_{1.875}$Ba$_{0.125}$CuO$_4$ and La$_{1.875}$Ba$_{0.075}$Sr$_{0.05}$CuO$_4$. *Phys. Rev. B* **70**, 104517 (2004).

[8] Doiron-Leyraud, N., *et al.,* Quantum oscillations and the Fermi surface in an underdoped high-$T_c$ superconductor. *Nature* **447**, 565-568 (2007).

[9] LeBoeuf, D., *et al*., Electron pockets in the Fermi surface of hole-doped high-$T_c$ superconductors. *Nature* **450**, 533 (2007).

[10] Laliberté*,* F., *et al.,* Fermi-surface reconstruction by stripe order in cuprate superconductors. *Nature Communications* **2**, 432 (2011).

[11] Liang, R., *et al.,* Evaluation of CuO$_2$ plane hole doping in YBa$_2$Cu$_3$O$_{6+x}$ single crystals. *Phys. Rev. B* **73**, 180505 (2006).

[12] Wu, T., *et al.,* Magnetic-field-induced charge-stripe order in the high-temperature superconductor YBa$_2$Cu$_3$O$_y$. *Nature* **477**, 191-194 (2011).

[13] von Zimmermann, M., *et al.,* Oxygen-ordering superstructures in underdoped YBa$_2$Cu$_3$O$_{6+x}$ studied by hard x-ray diffraction. *Phys. Rev. B* **68**, 104515 (2003).

[14] Abbamonte, P., Charge modulations versus strain waves in resonant x-ray scattering. *Phys. Rev. B* **74**, 195113 (2006).

[15] Schmidt, A. R., *et al*., Electronic Structure of the Cuprate Superconducting and Pseudogap Phases from Spectroscopic Imaging. *New Journal of Physics* **13**, 065014 (2011).

[16] Hoffman, J. E., *et al*., A Four Unit Cell Periodic Pattern of Quasi-Particle States Surrounding Vortex Cores in Bi$_2$Sr$_2$CaCu$_2$O$_{8+\delta}$. *Science* **295**, 466 (2002).

[17] Wise, W. D., *et al,* Charge-density-wave origin of cuprate checkerboard visualized by scanning tunnelling microscopy. *Nature Physics* **4**, 696-699 (2008).

[18] Reznik, D., *et al.,* Electron-phonon coupling reflecting dynamic charge inhomogeneity in copper oxide superconductors. *Nature* **440**, 1170-1173 (2006).

[19] Fournier, D., *et al*., Loss of nodal quasiparticle integrity in underdoped YBa$_2$Cu$_3$O$_{6+x}$. *Nature Physics* **6**, 905–911 (2010).

[20] Carrington, A., and Yelland, E. A., Band-structure calculations of Fermi-surface pockets in ortho-II YBa$_2$Cu$_3$O$_{6.5}$, *Phys. Rev. B* **76**, 140508 (2007).

[21] Harrison, N., Near Doping-Independent Pocket Area from an Antinodal Fermi Surface Instability in Underdoped High Temperature Superconductors. *Phys. Rev. Lett.* **107**, 186408 (2011).

[22] Yao, H., Lee, D.-H., and Kivelson, S., Fermi-surface reconstruction in a smectic phase of a high-temperature superconductor. *Phys. Rev. B* **84**, 012507 (2011).

[23] Daou, R., *et al*., Broken rotational symmetry in the pseudogap phase of a high-$T_c$ superconductor. *Nature* **463**, 519-522 (2010).

[24] LeBoeuf, D., *et al*., Lifshitz critical point in the cuprate superconductor YBa$_2$Cu$_3$O$_y$ from high-field Hall effect measurements. *Phys. Rev. B* **83**, 054506 (2011).

[25] Chang J., *et al*., Nernst effect in the cuprate superconductor YBa$_2$Cu$_3$O$_y$: Broken rotational and translational symmetries. *Phys. Rev. B* **84**, 014507 (2011).

[26] Chang, J., *et al.*, Nernst and Seebeck coefficients of the cuprate superconductor YBa$_2$Cu$_3$O$_{6.67}$: a study of Fermi surface reconstruction. *Phys. Rev. Lett.* **104**, 057005 (2010).

[27] Wang, Y., *et al*., High field phase diagram of cuprates derived from the Nernst effect. *Phys. Rev. Lett.* **88**, 257003 (2002).

[28] Balseiro, C. A., and Falicov, L. M., Superconductivity and charge-density waves. *Phys. Rev. B* **20**, 4457 (1979).

[29] Gabovich, Alexander M. and Voitenko, Alexander I., Model for the coexistence of *d*-wave superconducting and charge-density-wave order parameters in high-temperature cuprate superconductors. *Phys. Rev. B* **80**, 224501 (2009)

[30] Haug, D., *et al*., Neutron scattering study of the magnetic phase diagram of underdoped YBa$_2$Cu$_3$O$_{6+x}$. *New J. Phys.* **12**, 105006 (2010).

[31] G. Ghiringhelli *et al.*, Long-range incommensurate charge fluctuations in (Y,Nd)Ba$_2$Cu$_3$O$_{6+x}$. To appear in Science (2012).

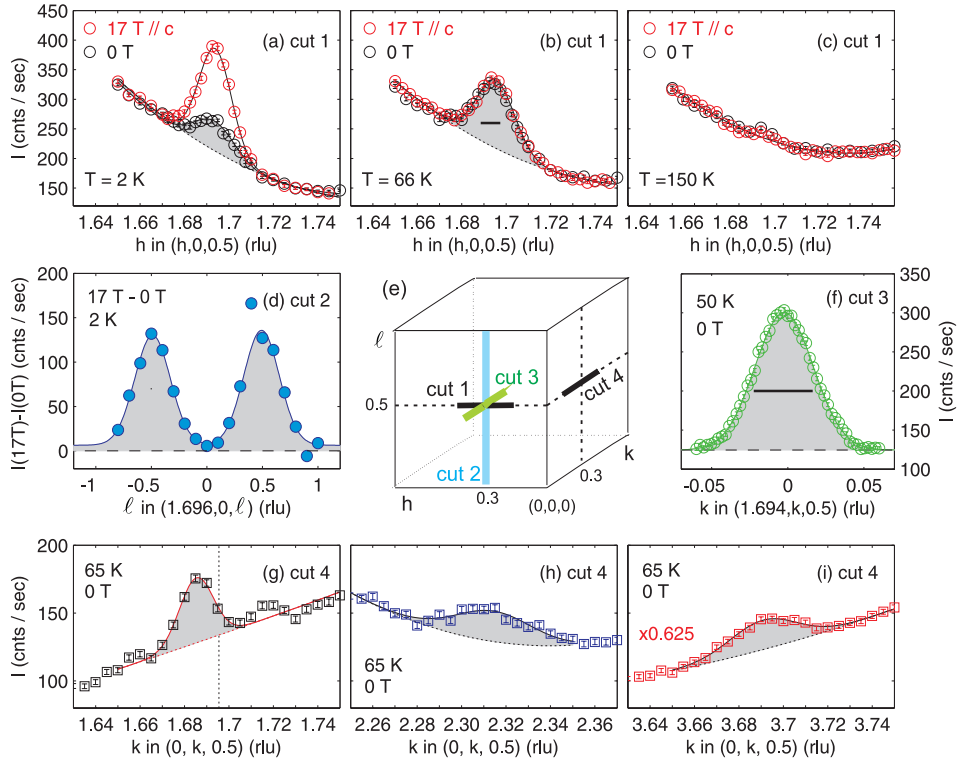

**Figure 1 | Incommensurate charge-density-wave order.**
Diffracted intensity in reciprocal space $\mathbf{Q}=(h,k,\ell)=h\mathbf{a}^*+k\mathbf{b}^*+\ell\mathbf{c}^*$ where $a^*=2\pi/a$, $b^*=2\pi/b$ and $c^*=2\pi/c$, with lattice parameters $a=3.81$ Å, $b=3.87$ Å (see Fig. S1), $c=11.72$ Å. Four different scans in reciprocal space, projected into first Brillouin zone, are shown schematically in Fig. 1(e). (a-c) Scans along ($h$, 0, 0.5) for temperatures and magnetic fields (applied along the crystal $\mathbf{c}$-direction) as indicated. An incommensurate lattice modulation, peaked at $(2-\delta_1, 0, 0)$ where $\delta_1 = 0.3045(2)$, emerges as the temperature is lowered below 135 K. The intensity of the satellite in (b) is of order $2\times10^{-6}$ weaker than the (2, 0, 0) reflection. This becomes field-dependent below the zero-field superconducting transition temperature $T_c = 67$ K. The full-width half-maximum (FWHM) instrumental resolution is shown by horizontal lines in (b) and (f). By deconvolving the resolution from the Gaussian fits to the data taken at 17 T and 2 K, an $h$-width of $\sigma_a = 6.4\times10^{-3}$ r.l.u. corresponding to a correlation length $\xi_a = 1/\sigma_a$ of $95\pm1$ Å was found (see Supplementary Information). (d) The field-induced signal $I(17T)-I(0T)$ at $T=2K$ is modulated along $(1.695, 0, \ell)$ and it peaks approximately at $\ell=\pm0.5$. (f) Scan along (1.695, $k$, 0.5). The poor resolution along the $\mathbf{k}$-direction did not allow accurate determination of the width along (1.695, $k$, 0.5) but we estimate a value of 0.01 r.l.u., comparable to that along ($h$, 0, 0.5), indicating similar coherence lengths along $\mathbf{a}$- and $\mathbf{b}$-axis directions. (g-i) Scans along (0,$k$,0.5). Incommensurate peaks are found in several Brillouin zones e.g. at positions $\mathbf{Q} = (0, 2\pm\delta_2, 0.5)$ and $(0, 4-\delta_2, 0.5)$ where $\delta_2=0.3146(7)$, see also Fig. S3. The vertical dashed line in (g) indicates $\delta_1$. The lattice modulation was fitted to a Gaussian function (solid lines in a-d,f-i) on a background (dashed lines) modelled by a second order polynomial. Error bars are determined by counting statistics.

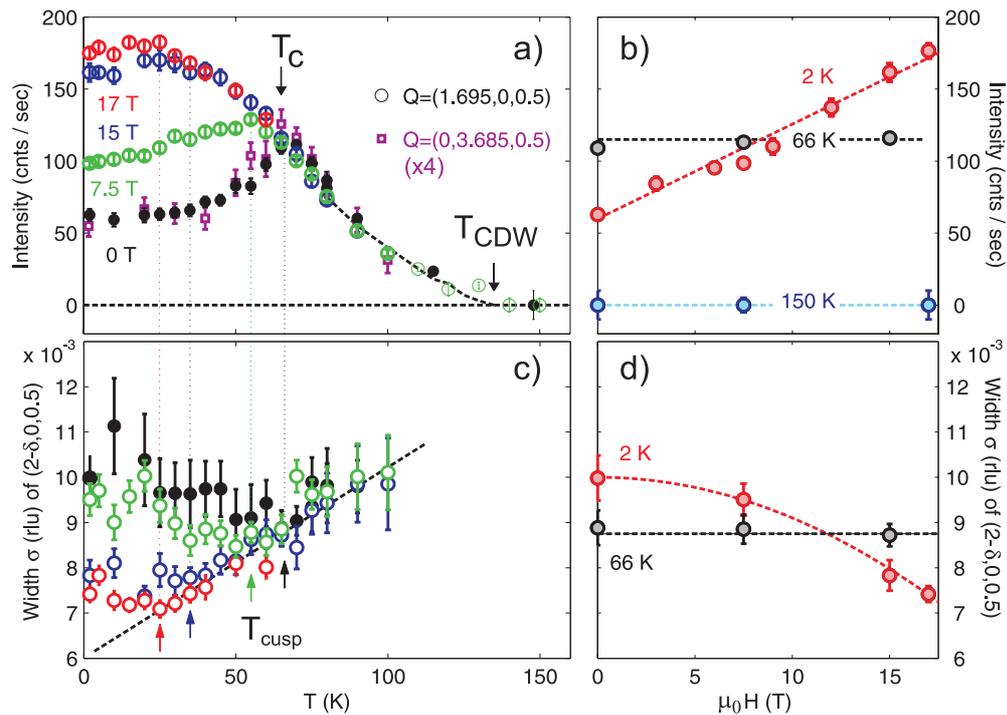

**Figure 2 | Competition between charge-density-wave order and superconductivity.**

(a) Temperature dependence of the peak intensity at (1.695, 0,0.5) (circles) and (0, 3.691, 0.5) (squares) for different applied magnetic fields. The square data points have been multiplied by a factor four. In the normal state, there is a smooth onset of the CDW order. In absence of an applied magnetic field there is a decrease in the peak intensity below $T_c$. This trend can be reversed by the application of a magnetic field. (b) Magnetic field dependence of the lattice modulation peak intensity at (1.695,0,0.5) for different temperatures. At $T=2K$, the peak intensity grows approximately linearly with magnetic field up to the highest applied field. (c-d) Gaussian linewidth of the (1.695,0,0.5) CDW modulation plotted versus temperature and field. The raw linewidth, including a contribution from the instrumental resolution, is field independent in the normal state ($T>T_c$). In contrast, the CDW order becomes more coherent below $T_c$, once a magnetic field is applied. This effect ceases once the amplitude starts to be suppressed due to competition with superconductivity. The vertical dashed lines in (a) and (c) illustrate the connection between these two features of the data that define the $T_{cusp}$ temperature scale. All other lines are guides to the eye. Error bars are the uncertainty of fitting parameters.

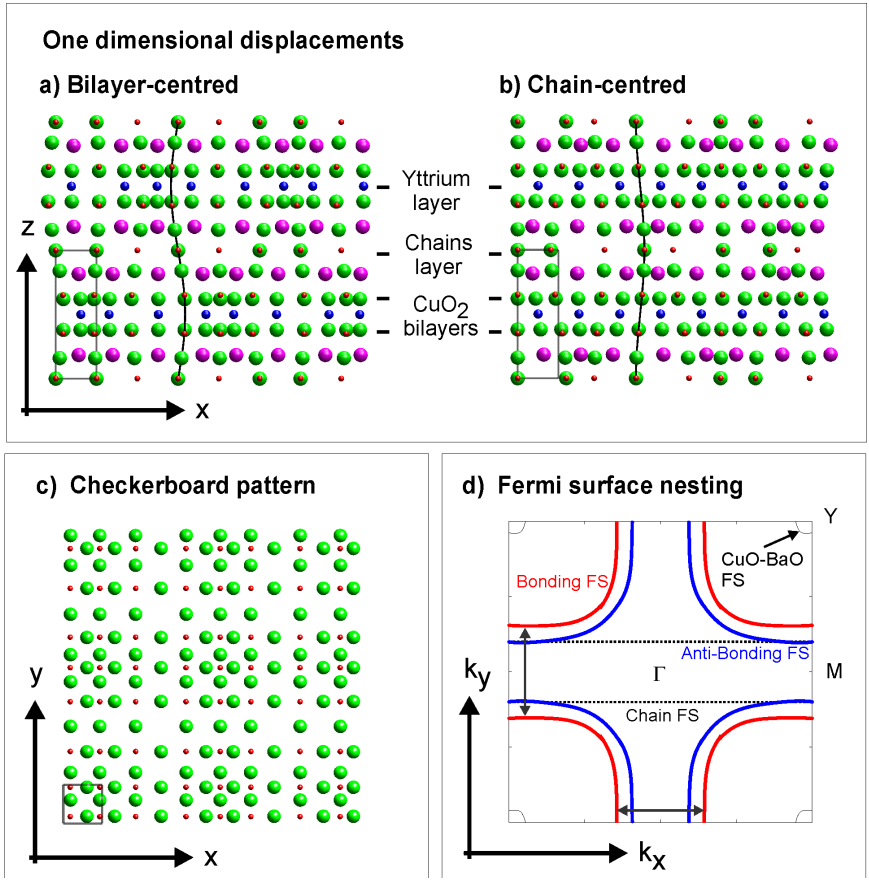

**Figure 3 | Real space and reciprocal space pictures.**
(a-b) One dimensional atomic displacements (exaggerated for clarity - the estimated amplitude is ~0.003 Å, see Supplementary Information) in the unit cell of YBCO ortho-VIII. The positions of yttrium (blue), barium (purple), copper (red) and oxygen (green) atoms are displayed as a function of *x* (Cu-O bond direction) and *z* perpendicular to the $CuO_2$ planes. The structure in (a) has maximal displacement on the yttrium site and the displacement is the same in the two bilayer $CuO_2$ planes. In panel (b), the structure has maximal displacement at CuO-chain positions and the displacement is opposite within the bilayer planes. Notice that the displacements in neighboring bilayers are in anti-phase and the CuO chains are modulated with the ortho-VIII structure. (c) Modulations along both Cu-O bond directions (*x* and *y*) lead to a checkerboard structure. (d) Schematics of the Fermi surface of YBCO (based on Ref. 20 – see Supplementary Information). It consists of bonding (red) and anti-bonding (blue) sheets originating from the $CuO_2$ bilayer planes, one-dimensional open sheets from the oxygen chains and small pockets around the Y-point from that CuO-BaO layers. Except for the last mentioned, these Fermi surface sheets have been observed by angle resolved photoemission spectroscopy measurements[19] on $YBa_2Cu_3O_y$. Coupling of the bonding states (indicated by vertical and horizontal arrows) favours the bilayer-centred structure, although this is in competition with intra-bilayer Coulomb effects [see Supplementary Information]. The arrows connecting the bonding bands near the zone boundary are compatible with the CDW vectors $q_1=(\delta_1, 0, 0.5)$ and $q_2=(0, \delta_2, 0.5)$ if the Fermi surface has a small dispersion along $k_z$.

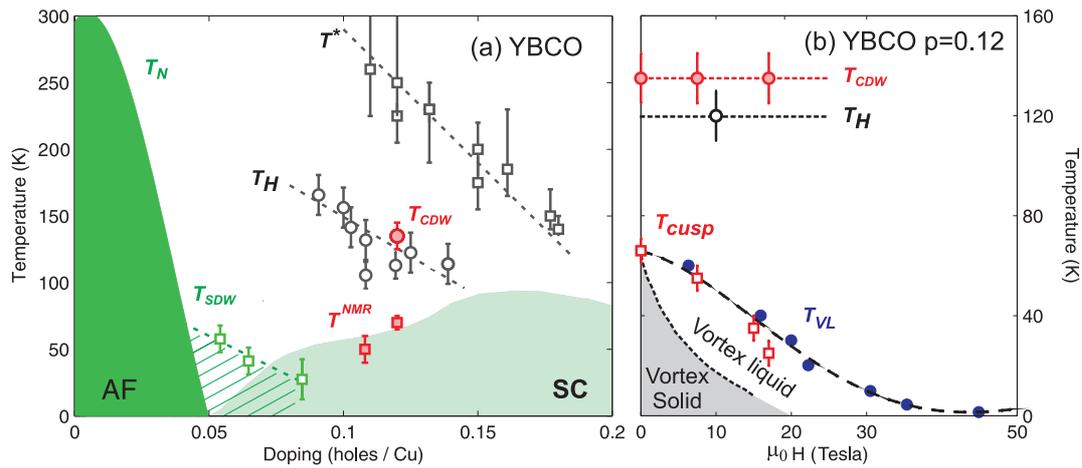

**Figure 4 | Phase diagram of $YBa_2Cu_3O_{7-x}$**

(a) Doping dependence of the antiferromagnetic ordering temperature $T_N$, incommensurate spin-density wave order $T_{SDW}$[30], the pseudogap temperature $T^*$[23,24] and the superconducting temperature $T_c$. Below temperature scale $T_H$[24,25], a larger and negative Hall coefficient was observed[24] and interpreted in terms of a Fermi surface reconstruction. Our X-ray diffraction experiments show that in YBCO $p$=0.12 incommensurate CDW order spontaneously breaks the crystal translational symmetry at a temperature $T_{CDW}$ that is twice as large as $T_c$. $T_{CDW}$ is also much larger than $T^{NMR}$, the temperature scale below which NMR observes field-induced charge order[12]. (b) Field dependence of $T_{CDW}$ (filled red circles) and $T_{cusp}$ (open squares), the temperature below which the CDW is suppressed by superconductivity, compared with $T_H$ (open black circle) and $T_{VL}$ (filled blue circles), the temperature where the vortex liquid state forms[24].